\documentclass[journal=jpccck,manuscript=article]{achemso}

\usepackage[version=3]{mhchem} 
\usepackage{graphicx}

\author{J. A. McLeod$^{1\dagger}$}
\email{john.mcleod@usask.ca}
\author{D. W. Boukhvalov$^2$}
\author{D. A. Zatsepin$^{3,4}$}
\author{R. J. Green$^{1\ddagger}$}
\author{B. Leedahl$^1$}
\author{L. Cui$^5$}
\author{E. Z. Kurmaev$^3$}
\author{I. S. Zhidkov$^{3,4}$}
\author{L. D. Finkelstein$^3$}
\author{N. V. Gavrilov$^6$}
\author{S. O. Cholakh$^4$}
\author{A. Moewes$^1$}

\affiliation{$^1$ Department of Physics and Engineering Physics, University of Saskatchewan, 116 Science Place, Saskatoon, Saskatchewan S7N 5E2, Canada\\
$^2$ School of Computational Sciences, Korea Institute for Advanced Study (KIAS) Hoegiro 87, Dongdaemun-Gu, Seoul, 130-722, Korean Republic\\
$^3$ Institute of Metal Physics, Russian Academy of Sciences--Ural Division,
620990 Yekaterinburg, Russia\\
$^4$ Ural Federal University, 620002 Yekaterinburg, Russia\\
$^5$ Shenzhen Graduate School, Harbin Institute of Technology, Shenzhen 518055, PR China\\
$^6$ Institute of Electrophysics, Russian Academy of Sciences-Ural Division, 620016 Yekaterinburg, Russia\\
$^\dagger$ (Current Address:) College of Nano Science and Technology, Soochow University, 199 Ren-Ai Rd., Suzhou Industrial Park, Suzhou, Jiangsu, 215123, China [jmcleod@suda.edu.cn]\\
$^\ddagger$ (Current Address:) Department of Physics and Astronomy, University of British Columbia, 6224 Agricultural Road, Vancouver, British Columbia V6T 1Z1, Canada}

\title[Fe-doped ZnO]
{Local Structure of Fe Impurity Atoms in ZnO: Bulk versus Surface}

\begin{document}
\begin{abstract}
By studying Fe-doped ZnO pellets and thin films with various x-ray spectroscopic techniques, and complementing this with density functional theory calculations, we find that Fe-doping in bulk ZnO induces isovalent (and isostructural) cation substitution (Fe$^{2+} \rightarrow$ Zn$^{2+}$). In contrast to this, Fe-doping near the surface produces both isovalent and heterovalent substitution (Fe$^{3+} \rightarrow$ Zn$^{2+}$). The calculations performed herein suggest that the most likely defect structure is the single or double substitution of Zn with Fe, although, if additional oxygen is available, then Fe substitution with interstitial oxygen is even more energetically favourable. Furthermore, it is found that ferromagnetic states are energetically unfavourable, and ferromagnetic ordering is likely to be realized only through the formation of a secondary phase (i.e. ZnFe$_2$O$_4$), or codoping with Cu.
\end{abstract}

\section{Introduction}
Over the last decade there has been a great deal of research in doping of the binary oxides ZnO and TiO$_2$ with a small percentage (in terms of total number of atoms) of 3$d$ transition metal ions. This research was driven by several reasons: among them the search for ferromagnetism above 300 K in a dilute magnetic semiconductor,~\cite{c1, xu_08} and the possibility of band gap engineering a semiconductor to act as a photocatalyst for water splitting,~\cite{c2,c3,c4} and the search for a new transparent conducting oxide for photovoltaics.~\cite{lee_10,raj_13} In both cases, however, the experimental results were strongly dependent on the preparation conditions, and great discrepancies were observed between thin film and bulk materials even in cases of identical composition (see, for example, Refs.~\cite{c5,c6} and references therein). In fact, while early theoretical treatment suggested that room-temperature ferromagnetic order should spontaneously arise in transition-metal doped ZnO materials,~\cite{sato_00} subsequent experiments failed to detect any ferromagnetism over large temperature ranges.~\cite{jin_01} Later experiments found room-temperature ferromagnetic order even in undoped ZnO, but only in thin films or nanocrystals,~\cite{inamdar_11} but the ferromagnetism was attributed to Zn vacancies, and the ferromagnetic order may or may not be stabilized by Fe-doping.~\cite{hong_07,karmakar_07}

The general problem when doping binary oxides is that both isovalent and heterovalent cation substitutions are possible, and this leads to the formation of a wide variety of structural defects including vacancies, interstitial ions, and precipitates.~\cite{c21} Furthermore, predicting the properties of doped binary oxides by purely theoretical means is limited due to the lack of a model properly describing the true configurations of the structural defects. This is largely due to the fact that small scale structures are not easily detected by conventional methods --- especially in thin films~\cite{c7} --- so computational approaches must consider a rather exhaustive range of trial geometries for defects.

In this work Fe-doped (with an Fe ion fluence of about 10$^{17}$ cm$^{-2}$) pellet and thin film ZnO are examined using a variety of x-ray spectroscopies and density functional theory (DFT) calculations. The former provides site- and symmetry-specific probing of the occupied and unoccupied states, while the latter provides specific details on the formation and couplings between different defect sites, and aid in interpreting the fine structure of the x-ray spectra. In addition to providing an empirical probe of the electronic structure over a broad energy range (i.e. the entire valence or conduction bands) local to an Fe atom, x-ray spectroscopy is also very sensitive to the valency of these atoms.~\cite{degroot_05} X-ray spectroscopy measurements can therefore be used to test the accuracy of the electronic structures calculated using DFT with a particular choice of pseudopotentials and exchange-correlation functional, and justify using these DFT calculations to predict other material properties.

\section{Experimental and Theoretical Methods}
\subsection{Sample Preparation}
Pellets of polycrystalline ZnO were made from a 99.999\% pure Zn target sputtered under a mixed plasma of Ar (99.999\%) and O$_2$ (99.999\%). The distance between the target and the substrate was 8.0 cm, the base pressure was $3 \times 10^{−4}$ Pa, and the flow rates of Ar (10 sccm) and O$_2$ (20 sccm) were controlled by mass flow controllers. The polycrystalline pellets were around 10 mm in diameter and 1 mm to 3 mm thick.

To deposit the thin films, a sapphire substrate (0001) was ultrasonically cleaned in acetone and alcohol for 10 minutes, then rinsed in deionized water, and finally dried in N$_2$. The sapphire substrates were held at a temperature of 250 $^\circ$C for 90 minutes during deposition, and the deposition was carried out at a working pressure of 2 Pa after pre-sputtering with Ar for 10 minutes. When the chamber pressure was stabilized, the radio frequency generator was set to 100 W. The growth rate of ZnO thin films was 3.4 nm/min and the typical thin film thickness was 302 nm. The polycrystalline ZnO samples had a hexagonal structure with lattice parameters $a = 3.250$ \AA~ and $c = 5.207$ \AA; XRD of these films showed there was a strong preference for the $c$ axis to be aligned vertically from the substrate surface. More details about sample preparation are available in our previous paper.~\cite{c8}

Commercial powders of ZnO, FeO, and Fe$_2$O$_3$ (Alfa Aesar, 99.99\% purity) were also acquired for reference measurements.

\subsection{Ion Implantation}
Both the pellet and thin film ZnO samples were implanted with Fe ions under vacuum (the residual pressure was $3\times10^{-3}$ Pa). A \mbox{30 keV} ion beam was generated by the source based on a cathodic vacuum arc. The arc was initiated with an auxiliary discharge in argon; this caused the pressure in the chamber during ion irradiation to increase to \mbox{$1.5\times10^{-2}$ Pa}. A current density of 0.7 mA/cm$^2$ was then delivered in pulse mode with a repetition rate of \mbox{25 Hz} and a pulse duration of 0.4 ms. After 38 minutes of exposure, the sample had an ion fluence of \mbox{$10^{17}$ cm$^{-2}$}. The samples were placed on a massive water-cooled collector during exposure. The initial temperature of the samples prior to irradiation was 20 $^\circ$C, and after implantation the samples were allowed to cool under vacuum for 20 minutes. Based on our stopping and range of ions in matter (SRIM) calculations,~\cite{srim_ref} the ion implantation has a peak density of $\sim 5.5 \times 10^{22}$ atoms/cm$^3$ at a depth of $\sim 150$~\AA, and an average density of $\sim 1.6\times 10^{22}$ atoms/cm$^3$ to a maximum depth of $\sim 650$~\AA. Finally, we note that the concentration of Fe atoms is perhaps too large to be considered ``doping'' in the sense commonly used in semiconductor physics; however we will persist in using the term ``doping'' in this manuscript and the reader is welcome to interpret this term as meaning ``chemical doping'' or even ``chemical substitution''.

\subsection{XPS Measurements}
\begin{figure}
\begin{center}
\includegraphics[width=3in]{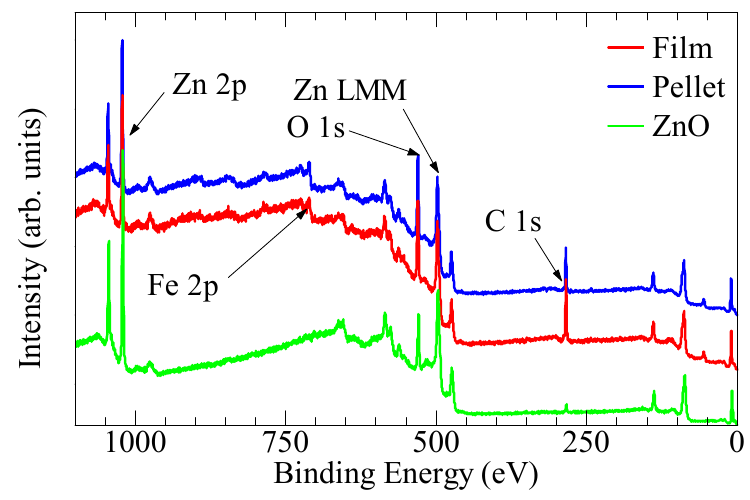}
\caption{XPS survey scans of Fe-doped pellet and thin film ZnO, as well as an undoped reference. The significant features are labeled. (Color online.)}
\label{fig:xps_survey}
\end{center}
\end{figure}

X-ray photoelectron spectroscopy (XPS) core-level and valence-band measurements were made using a PHI XPS Versaprobe 5000 spectrometer (ULVAC-Physical Electronics, USA) based on a classic x-ray optic scheme with a hemispherical quartz monochromator and an energy analyzer working in the range of binding energies from 0 to 1500 eV. This apparatus uses electrostatic focusing and magnetic screening to achieve an energy resolution of $\Delta$E $\leq$ 0.5 eV for Al K$\alpha$ radiation (1486.6 eV). The samples were introduced to vacuum (10$^{-7}$ Pa) for 24 hours prior to measurement, and only samples whose surfaces were free from micro-impurities were measured and reported herein. The XPS spectra were recorded using Al $K\alpha$ x-ray emission; the spot size was 100 $\mu$m, and the x-ray power load delivered to the sample was less than 25 W. Typical signal to noise ratios were greater than 10000:3. Finally, the spectra were processed using ULVAC-PHI MultiPak Software 9.3 and the residual background was removed using the Tougard method.~\cite{c9} The XPS spectra were calibrated using a reference energy of 285.0 eV for the carbon 1$s$ core level.\cite{c10} The survey scans, shown in Figure~\ref{fig:xps_survey}, reveal a relatively low carbon content and, more importantly, show no indication of metal impurities other than Fe. 

\subsection{XES and XAS Measurements}
X-ray emission spectroscopy (XES) and x-ray absorption spectroscopy (XAS) measurements at the Fe $L$-edge and O $K$-edge were performed using Beamline 8.0.1~\cite{jia_95} at the Advanced Light Source (ALS) at the Lawrence Berkeley National Laboratory. The beamline uses a Rowland circle type grating spectrometer with spherical gratings. Photons emitted from the sample were detected at an angle of 90$^\circ$ to the incident photons, and the incident photons were 30$^\circ$ to the sample surface normal with a linear polarization in the horizontal scattering plane. The measurements were performed in a vacuum chamber at $10^{-5}$ Pa, the XAS resolving power ($E/\Delta E$ was about $2\times10^3$, while the XES resolving power was about 10$^3$.

The Fe $L$-edge XAS spectra were calibrated using a reference energy of 708.4 eV for the first peak in the $L_3$ absorption edge and a reference splitting of 13.5 eV between the $L_3$ and $L_2$ absorption lines of metallic iron; the XES spectra were then calibrated with respect to the elastic scattering peaks from incident x-rays with energies resonant with the $L_2$ and $L_3$ absorption lines. The O $K$-edge XES and XAS spectra were calibrated using a reference energy of 526.0 eV and 532.7 eV for the O $K$ emission line and absorption edge of Bi$_4$Ge$_3$O$_{12}$, respectively.  High-resolution XAS measurements in both bulk sensitive total fluorescence yield (TFY) and surface sensitive total electron yield (TEY) modes were later performed at the Fe $L$-edge and O $K$-edge using the SGM beamline at the Canadian Light Source,~\cite{regier_07} these spectra were then calibrated in same way as the aforementioned XAS measurements. We point out that the absolute scale of these measurements is not strictly relevant for our discussion herein.

\subsection{DFT Calculations}
The density-functional theory (DFT) calculations were performed using the SIESTA pseudopotential code,~\cite{SIESTA_1, SIESTA_2} in a manner similar to that previously used in related studies of impurities in semiconductors.~\cite{c13} All calculations were performed using the Perdew-Burke-Ernzerhof variant of the generalized gradient approximation (GGA-PBE)~\cite{c14a} for the exchange-correlation potential. A full optimization of the atomic positions was performed, during which the electronic ground state was consistently found using norm-conserving pseudopotentials for cores and a \mbox{double-$\zeta$} plus polarization basis of the localized orbitals for Fe, Zn, and O. The forces and total energies were optimized with an accuracy of \mbox{0.04 eV/\AA} and \mbox{1 meV}, respectively. For the atomic structure calculations, a Zn pseudopotential with Zn 3$d$ electrons treated as localized core states was employed. All calculations were performed with an energy mesh cut-off of 360 Ry and a $k$-point mesh of $6\times 6\times 6$ using the Monkhorst--Park scheme.~\cite{c15a} 

\begin{figure}
\begin{center}
\includegraphics[width=3in]{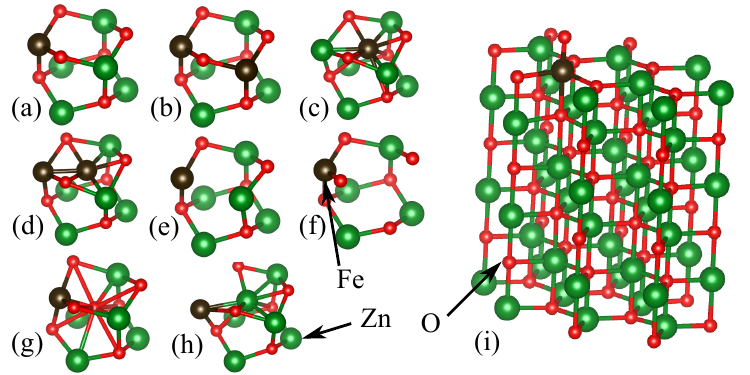}
\caption{Optimized structures of bulk ZnO in the vicinity of various defects such as: (a) single iron substitution ($S$), (b) double iron substitution ($2S$), (c) interstitial iron ($I$), (d) single iron substitution and interstitial iron ($S+I$), (e) single iron substitution and an oxygen vacancy ($S+V_O$), (f) single iron substutition and a zinc vacancy ($S+V_{Zn}$), (g) single iron substitution with interstitial oxygen ($S+I_{O}$), and (h) single iron substitution with interstitial zinc ($S+I_{Zn}$). The optimized structure of a slab used for modelling surface iron substitutions passivated by oxygen is shown in (i). (Color online.)}
\label{fig:structures}
\end{center}
\end{figure}

The formation energy ($E_\mathrm{form}$) calculations were performed using the standard method described in detail in Ref.~\cite{c13}. Exchange interactions between the clusters were calculated by the formula $J = (E_{AFM} - E_{FM})/ 2 S^2$, where $E_{FM}$ and $E_{AFM}$ are the total energies corresponding to parallel and antiparallel orientation of spins on the remote clusters, respectively, and $S$ is the total spin of the Fe atoms. To understand the nature of the ferromagnetism in the materials, different combinations of structural defects induced by Fe ion implantation were calculated in the ZnO matrix. As a host for the studied defects, a ZnO supercell consisting of 108 atoms was used. We then performed calculations for various combinations of substitutional ($S$) and interstitial ($I$) Fe impurities (see Fig. \ref{fig:structures}, (a)--(h)). Point defects such as oxygen and zinc vacancies, and interstitial atoms (further denoted as $V_{O}$, $V_{Zn}$, $I_{O}$, and $I_{Zn}$) have also been considered herein. To model an iron impurity on the ZnO surface, a six layer slab was used (see Fig. \ref{fig:structures}, (i)). For more details of our computational method see our previous work on modelling transition metal and aluminum impurities in semiconductors.~\cite{c13}

We would like to note that our choice of using GGA-PBE, rather than a more complex approach such as using a hybrid functional or on-site Hubbard $U$, was driven by the facts that GGA-PBE is computationally much simpler than the other approaches, GGA-PBE adequately reproduces the electronic structure of TM-doped DMS systems,~\cite{c13} and the more complex approaches provide virtually the same results in terms of defect structures and exchange interactions.~\cite{garcia_12,sato_10} More detailed theoretical analyses of defect formation (such as oxygen vacancies) in ZnO can be found in the literature,~\cite{lany_08,lany_10} while the formation energies reported herein may not be as accurate as those determined with more sophisticated methods, the trends reported herein are essentially the same as those found elsewhere in the literature. Likewise both our DFT calculations and spectroscopy measurements suggest that the band gap of ZnO is reduced after Fe-doping, however neither technique is suited to accurately determining what the band gap of Fe-doped ZnO is (GGA is well known to underestimate band gaps,~\cite{dufek_94} and the core hole shift in the XAS spectrum of Fe-doped ZnO is unknown).

\subsection{Multiplet Calculations}
The Fe $L$-edge RXES and XAS spectra were simulated using ligand field multiplet calculations.~\cite{c15b,laan_88} For simplicity, the local symmetry of an Fe substitution was treated as tetrahedral (the true point group of Zn in hexagonal ZnO is $C_{6v}$, but this is similar to tetrahedral). The 10$Dq$ value (this value represents the splitting between the $e_g$ and $t_{2g}$ bands) used in the calculations was \mbox{-0.5 eV}; this was chosen by comparing the calculated spectra to the experimental measurements. The calculated spectra were broadened via a convolution with a Lorentzian profile (to incorporate the influence of life-time broadening) and a Gaussian profile (to incorporate the influence of instrumental broadening).

\section{Results and Discussion}

\subsection{Spectroscopy Measurements}
X-ray spectroscopy provides a means of directly probing atomic and symmetry-selective states of a system, and is therefore a valuable tool in studying the transition metal doping of ZnO. XPS spectroscopy directly probes occupied states and can therefore be used to profile the total (occupied) density of states (DOS) in the valence band. Secondly, due to the final-state interaction between a core-hole and the valence band, core-level XPS (in the present case, the direct excitation of Fe 2$p_{1/2}$ or 2$p_{3/2}$ states) can provide information about the chemical environment of the absorbing atom.~\cite{c15b,lee_91} In XAS spectroscopy, an x-ray is absorbed by a core level electron, this electron is consequently  excited to an (unoccupied) conduction band state after the absorption of an x-ray; this transition is governed by dipole selection rules. Similarly, XES spectroscopy involves a dipole selection-driven transition from an occupied valence state to an unoccupied core level combined with the emission of an x-ray. XAS and XES can therefore probe the local partial DOS of the conduction and valence bands, respectively. Finally, when an XES spectrum is taken when the excitation energy is resonant with an absorption feature, resonant inelastic x-ray scattering (RIXS) can occur. This changes the shape of the XES spectrum when compared to the non-resonant case, and these resonant excitations offer a local probe of magnetic, multiplet, and charge-transfer transitions.~\cite{kotani_01}

\begin{figure}
\begin{center}
\includegraphics[width=3in]{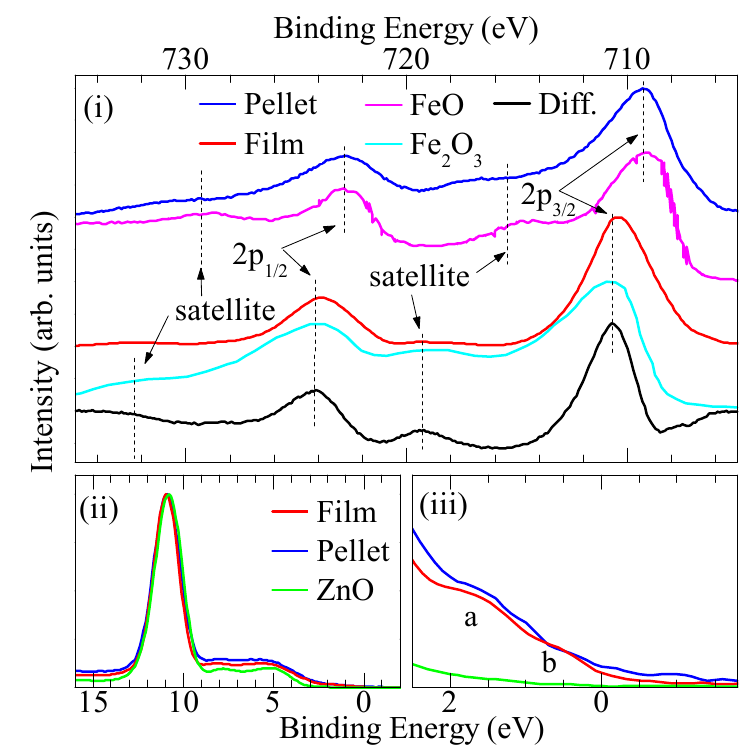}
\caption{XPS measurements of Fe-doped ZnO: (i) The Fe 2$p$ XPS spectra from a Fe-doped ZnO pellet and thin film, FeO,~\cite{c14b}, Fe$_2$O$_3$,~\cite{c15b} and a ``difference spectrum'' obtained by subtracting 70\% of the pellet spectrum from the thin film spectrum. (ii) The XPS VB spectra of Fe-doped pellet, thin film, and pure ZnO. (iii) An enlarged view of the VB spectra near the band gap, showing the mid-gap valence structures in the Fe-doped samples. (Color online.)}
\label{fig:xps}
\end{center}
\end{figure}

The XPS core and valence level spectra for Fe-doped and pure ZnO are shown in Fig. \ref{fig:xps}. The local structure of the iron impurities is studied by XPS Fe 2$p$-core level spectra, as shown in Fig. \ref{fig:xps} (i). The 2$p_{1/2}$ and 2$p_{3/2}$ peaks are clearly visible, and each has an accompanying satellite line. It is clear that the Fe-doped ZnO XPS spectrum in the pellet is quite close to that of FeO,~\cite{lee_91} indicating that the Fe-doping in the pellet is primarily due to isovalent substitution (i.e. Fe$^{2+}$). On the other hand, the spectrum from the Fe-doped ZnO thin films resembles something in between the Fe$^{2+}$ in FeO and the Fe$^{3+}$ in Fe$_2$O$_3$. This suggests that the iron in the Fe-doped ZnO thin film is a mixture of both Fe$^{2+}$ and Fe$^{3+}$, indicating that likely both isovalent and heterovalent substitutions occur. This finding is in agreement with previous studies.~\cite{karmakar_07} The shapes and satellite lines in the spectra from FeO and Fe$_2$O$_3$ are well known in the literature.~\cite{bocquet_92,brundle_77,c14b,c15b}

By subtracting 70\% of the pellet Fe 2\textit{p} XPS spectrum from the thin film Fe 2\textit{p} XPS spectrum a ``difference spectrum'' is produced and shown in Fig. \ref{fig:xps} (ii); the features in this spectrum are energetically aligned with those of Fe$_2$O$_3$. This qualitative treatment suggests that the Fe-doping in thin film ZnO is approximately 30\% Fe$^{3+}$ and 70\% Fe$^{2+}$, to within the probe depth of XPS (some justification and discussion of these values may be found at the end of this section). Charge compensation for Fe$^{3+}$ substitution can be realized by Zn vacancies, or interstitial oxygen. We should add that from these measurements we cannot entirely rule out the presence of trace amounts of Fe$^{3+}$ on the surface of the ZnO pellets (for example, one could claim that the 2$p$ peaks in the XPS spectrum from the pellet are slightly higher in binding energy than those in the spectrum from FeO, as shown in Fig. \ref{fig:xps}(i)), however by comparing the shape of the Fe 2$p$ XPS spectra of FeO, Fe$_2$O$_3$, and Fe-doped ZnO pellets, we estimate that less than 10\% of the Fe sites probed in Fe-doped ZnO pellet are Fe$^{3+}$. However we stress that the presence of any Fe$^{3+}$ on the surface of these pellets would increase the estimated concentration of Fe$^{3+}$ in the Fe-doped thin film ZnO discussed above.

The XPS valence band spectra of pure and Fe-doped ZnO are shown in Fig. \ref{fig:xps} (ii). The Zn 3$d^{10}$ shell is located at $\sim$10.9 eV, but the majority of the valence band (between 5 and 9 eV) is due to the bonding of O 2$p$ states. Fe-doping introduces states near the Fermi level (see Fig. \ref{fig:xps} (iii)), that is, at energies above where the top of the valence band would be in pure ZnO. A mid-gap feature due to Fe 3$d$ states appears at $\sim$1.5 eV (labelled $a$ in Fig. \ref{fig:xps} (iii)) in both the ZnO pellet and thin film, while a secondary feature appears at $\sim$0.5 eV (labelled $b$ in Fig. \ref{fig:xps} (iii)) in the thin film sample. The former feature is likely a consequence of Fe$^{2+}$ substitution, and the appearance of this feature inside the nominal band gap of ZnO is consistent with the general strategy of band gap engineering of ZnO and TiO$_2$-based photocatalysts wherein transition metal $d^n$ states substitute the host $d^{10}$ or $d^{0}$ states, respectively.~\cite{c17} On the other hand, the latter feature is likely a consequence of Fe$^{3+}$ substitution, and the appearance of these states close to the Fermi level is usually considered the main reason for the appearance of ferromagnetism in ZnO:Fe thin films.~\cite{c18}

\begin{figure}
\begin{center}
\includegraphics[width=3in]{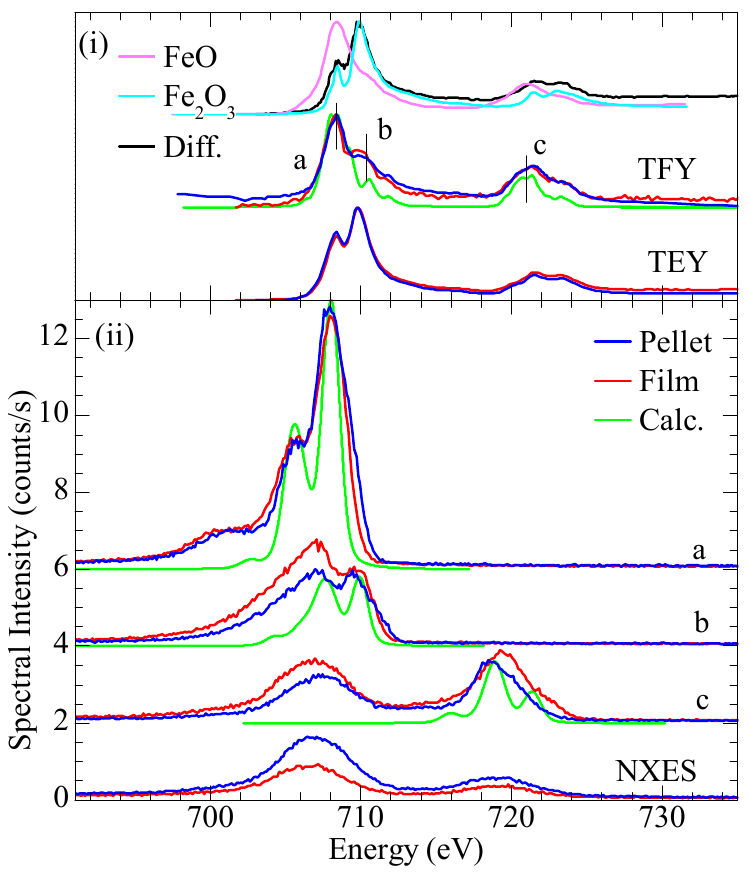}
\caption{Fe $L$ XAS and XES measurements of Fe-doped ZnO: (i) The XAS spectra were acquired in both surface-sensitive TEY and bulk-sensitive TFY modes, here they are compared with ligand-field multiplet calculations for Fe$^{2+}$, and reference spectra from FeO and Fe$_2$O$_3$. The letters $a$, $b$, $c$ denote the resonant excitation energies for the XES spectra. The normalized difference between the Fe-doped thin film TEY spectrum and 70\% of the Fe-doped pellet TEY spectrum is shown in black. (ii) The resonantly-excited XES spectra  ($a$, $b$, and $c$) and ligand-field multiplet calculations of Fe$^{2+}$. The ``NXES'' spectrum was acquired with an excitation energy of 750 eV, far above resonance, and is therefore not comparable to ligand-field multiplet calculations. (Color online.)}
\label{fig:fe_rixs}
\end{center}
\end{figure}

The XAS measurements of the Fe $L$-edge in the two Fe-doped ZnO samples are shown in Fig. \ref{fig:fe_rixs} (i). The bulk-sensitive TFY XAS spectra for both the pellet and thin film samples are effectively the same; this suggests that the Fe$^{3+}$ observed in the thin film is primarily at the surface. The TFY XAS is also in good agreement with the calculated spectrum of tetrahedrally-coordinated Fe$^{2+}$ with a 10$Dq$ splitting of -0.5 eV (see Fig. \ref{fig:fe_rixs} (i)). This provides additional confirmation that Fe$^{2+}$ simply substitutes for Zn in the wurtzite lattice, and does not significantly distort the structure. The shape of these spectra is also close to that from FeO, as shown in Fig. \ref{fig:fe_rixs} (i). The surface-sensitive TEY XAS is less straightforward, this is because the surface iron can adopt different local geometries, different ligand coordinations, and is more prone to oxidation. The shape of the TEY XAS spectra from both Fe-doped pellet and thin-film ZnO resemble a mixture of the spectra from both FeO and Fe$_2$O$_3$, as shown in Fig. \ref{fig:fe_rixs}(i). Since the XAS and RIXS measurements were performed some time after the XPS measurements, and the surface of these samples was not cleaved or sputtered prior to measurement, it is likely that prolonged exposure to air leads to further oxidation of the surfaces of both of these samples, consequently the TEY XAS observation of a more Fe$^{3+}$ rich surface compared to the XPS measurements is not surprising. Again, however, if 70\% of the TEY XAS spectrum from the Fe-doped ZnO pellet is subtracted from the TEY XAS spectrum of the Fe-doped ZnO thin film ZnO the resulting different is almost exactly the same shape as the Fe$_2$O$_3$ spectrum, as shown in Fig. \ref{fig:fe_rixs}(i).

The Fe $L$-edge in XES measurements are shown in Fig. \ref{fig:fe_rixs} (ii). Like the TFY XAS, the RIXS present in these spectra are in good agreement with the calculated RIXS of tetrahedrally-coordinated Fe$^{2+}$ with a 10$Dq$ splitting of -0.5 eV. Note that these calculations simply model the $dd$ multiplet transitions; the non-resonant emission [labelled ``NXES'' in Figure \ref{fig:fe_rixs} (ii)] and the charge transfer features at the low energy end of the spectrum were not taken into account in the calculations. The agreement between both the measured RIXS and XAS with the tetrahedrally-coordinated Fe$^{2+}$ calculations provides strong evidence that Fe-doping in ZnO provides Fe$^{2+}$ substitution in the Zn sites.

\begin{figure}
\begin{center}
\includegraphics[width=3in]{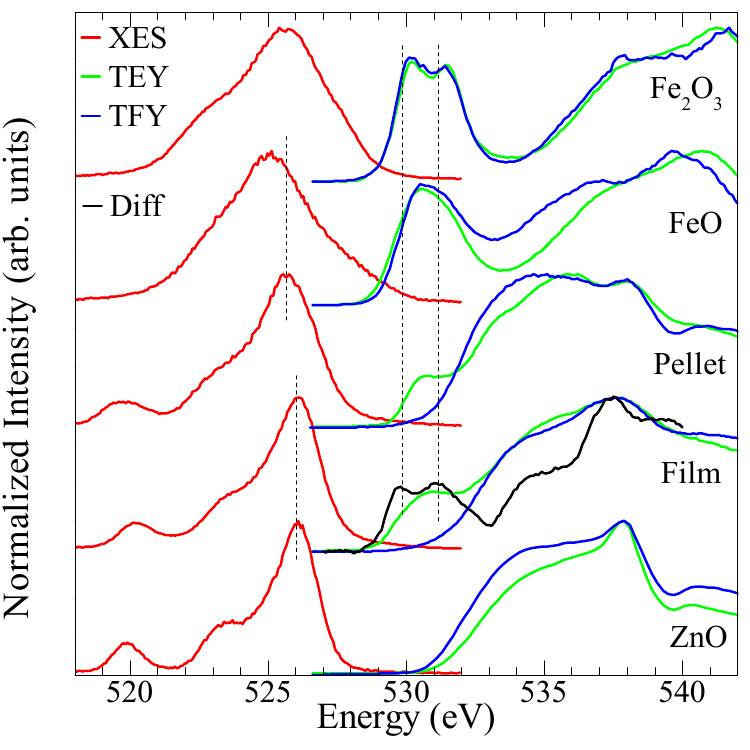}
\caption{O $K$ XES and XAS measurements of Fe-doped ZnO, pure ZnO, FeO, and Fe$_2$O$_3$. The XAS spectra were acquired in both surface-sensitive TEY and bulk-sensitive TFY modes as shown. The difference between Fe-doped ZnO thin film TEY spectrum and 70\% of the Fe-doped ZnO pellet TEY spectrum is shown in black, note the resemblance to Fe$_2$O$_3$. (Color online.)}
\label{fig:o_sxs}
\end{center}
\end{figure}

The O $K$-edge non-resonant XES and XAS measurements of pure and Fe-doped ZnO, as well as FeO and Fe$_2$O$_3$, are shown in Fig. \ref{fig:o_sxs}. The O $K$-edge XES of Fe-doped ZnO looks similar to that of pure ZnO, this is expected because the Fe-doping is relatively minimal. Because of the very different crystal structures, and the full Fe-coordination, the O $K$-edge XES of FeO and Fe$_2$O$_3$ is quite different than that of the Fe-doped ZnO samples. The peak of the O $K$ XES spectrum of the Fe-doped ZnO pellet is slightly lower in energy than the peak in the spectrum of pure ZnO, this can be attributed to the influence of Fe-doping since the peaks of the spectra of FeO and Fe$_2$O$_3$ also occur at lower energies in comparison with the spectrum of ZnO (see Fig. \ref{fig:o_sxs}).

Likewise, the O $K$-edge in XAS spectra, most notably in the bulk-sensitive TFY spectra, of the Fe-doped ZnO samples are quite close to that of pure ZnO. This is expected because the implantation depth is only around 10-60 nm, while the x-ray attenuation length at the O $K$-edge is a few hundred nm, and therefore the XES and TFY XAS are mostly probing pure polycrystalline ZnO. However there is a clear pre-edge feature in the surface-sensitive TEY spectra from both the Fe-doped ZnO pellet and thin film that is similar to the pre-edge feature in the O $K$-edge spectrum of FeO. Indeed, subtracting 70\% of the O $K$-edge TEY spectrum from the Fe-doped ZnO pellet from the O $K$-edge TEY spectrum of the thin film provides the ``difference spectrum'' in Fig. \ref{fig:o_sxs}, which has the characteristic two-peak pre-ege feature of Fe$_2$O$_3$. While we admit that this analysis is not precise, we have three separate measurements (Fe 2$p$-edge XPS, Fe \textit{L}$_{2,3}$ TEY XAS, and O $K$-edge XAS) both suggesting that the surface of the thin film has about 30\% Fe$^{3+}$ content (or at the Fe$^{3+}$ content is 30\% \textit{greater} in the thin film than in the pellet sample).

Finally, note that the relatively low energy of the Fe-related pre-edge features in the O $K$-edge XAS of Fe-doped ZnO indicates that Fe-doping reduces the band gap (at least at the surface), this was also suggested by the valence band XPS spectra. Because the O 1$s$ core hole has a very minor perturbative effect on the energy of the unoccupied states (i.e., the onset of the O $K$-edge XAS is very close to the true ground-state conduction band),~\cite{mcleod_12} the pre-edge features in the TEY spectra can be used to identify the bottom of the conduction band. Unfortunately, the Fe $L$-edge spectra, which provides the most direct probe of the electronic structure in the vicinity of the dopants, cannot be used to estimate the band gap because the Fe 2$p$ core hole significantly perturbs the onset of the Fe $L$-edge XAS from the true ground-state conduction band (in fact,the magnitude of this perturbation is one of the reasons why a ligand field calculation reproduces the XAS spectrum so well: the material-specific band 
structure is largely ``washed out'' and can be replaced by a single energy parameter).~\cite{mauchamp_09}

Before discussing our DFT calculations of these structures, we would like to briefly return to the subject of the ``difference spectra'' we have acquired from the Fe 2\textit{p} XPS, Fe \textit{L} TEY XAS, and O \textit{K} TEY XAS spectra of the pellet and film samples. In all cases, we subtracted 70\% of the spectrum from the pellet sample from the spectrum from the thin film sample, and concluded from the shape of the resulting ``difference spectrum'' compared to the corresponding spectrum of Fe$_2$O$_3$ that the thin film sample had at least 30\% more Fe$^{3+}$ near the surface than the pellet sample.

The value of 70\% scaling for the spectrum from the pellet sample was not chosen arbitrarily. If, for pellet and thin film spectra $y_{pellet}$ and $y_{film}$, we define the difference spectrum as $y_{\mathrm{Diff}} = a(y_{\mathrm{film}} - \alpha y_{\mathrm{pellet}}) + b$, for some renormalization constants $a$ and $b$, we can quantitatively examine the similarity between this ``difference spectrum'' and the appropriate spectrum from the Fe$_2$O$_3$ reference, $y_{\mathrm{Ref}}$, as the root-mean-square (RMS) error defined in Equation~\ref{eqn:rms} (where $E_i$ is the $i$th energy value in the discrete spectra).

\begin{equation}
\mathrm{RMS} = \sqrt{\frac{1}{N}\sum_i^N \left( y_{\mathrm{Ref}}(E_i) - y_{\mathrm{Diff}}(E_i)\right)^2}
\label{eqn:rms}
\end{equation}
The lower the RMS, the closer the ``difference spectrum'' is to the Fe$_2$O$_3$ reference. By choosing the renormalization constants $a$ and $b$ such that the RSS value is minimized for a particular choice of $\alpha$, we find that the best-fit value of $\alpha$ for the Fe 2\textit{p} XPS, Fe \textit{L} XAS, and O \textit{K} XAS are (to two significant figures) $\alpha = 0.46$, $\alpha = 0.67$, and $\alpha = 0.84$, respectively. The average of these values is $\alpha_{avg} = 0.7$, which gives some justification to our claim that the thin film has $1 - \alpha_{avg} = $ 30\% or so more Fe$^{3+}$ near the surface than the pellet sample.

\begin{figure}
\begin{center}
\includegraphics[width=3in]{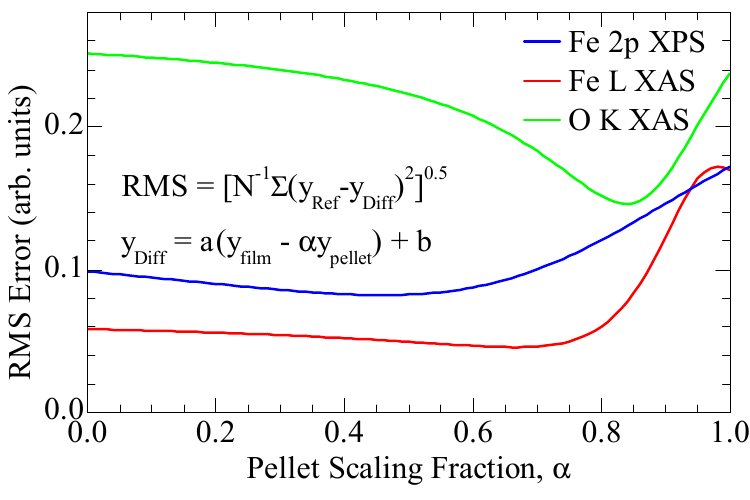}
\caption{The root-mean-square (RMS) errors for the Fe 2\textit{p} XPS, Fe \textit{L} XAS, or O \textit{K} XAS ``difference spectra'' and the reference Fe$_2$O$_3$ spectra. (Color online.)}
\label{fig:rms}
\end{center}
\end{figure}

We wish to stress that our claim of 30\% more Fe$^{3+}$ near the surface of the thin film sample than the pellet sample is still just a crude estimate, despite the justification for the value of 30\% given above. Indeed, the standard deviation of the best-fit $\alpha$  values suggests that we have $\alpha_{avg} = 0.7 \pm 0.2$, so the claim of 30\% more Fe$^{3+}$ is certainly not very accurate. For simplicity, we have neglected to take the relative sharpness of the minima for each best-fit $\alpha$ into account - certainly the relative flatness of the Fe 2\textit{p} XPS RMS error curve in Figure~\ref{fig:rms} suggests greater intrinsic uncertainty than there is in the relatively narrow minima in the O \textit{K} XAS RMS error curve. However since using the RMS error between a linearly-scaled ``difference spectrum'' and a reference standard is hardly the definitive method for accurately quantifying the relative concentrations of Fe$^{3+}$ and Fe$^{2+}$, we feel justified in claiming that there is 30\% more surface Fe$^{3+}$ in the thin film sample compared to the pellet sample as simply a semi-quantitative estimate, and rely on this discussion to emphasize the uncertainty of this approach. In other words, we feel confident in claiming that the surface of the thin film sample is richer in Fe$^{3+}$ compared to the surface of the pellet sample, but our claim that there is 30\% more Fe$^{3+}$ is just an attempt to quantify the previous statement, and the value of 30\% contains a considerable margin of error.
 
\subsection{DFT Calculations}
Our x-ray spectroscopy measurements provide a strong indication that Fe-doping in ZnO shows that Fe$^{2+}$ substitution occurs at Zn sites in the bulk of the sample; and that Fe$^{3+}$ substitution occurs near the surface (note that both XAS in TEY mode and XPS have a probe depth of only a few nanometres). We therefore have empirical data to test the validity of our DFT structural optimizations and electronic structure calculations, and can use these calculations to gain greater insight into the influence of Fe-doping in ZnO.

The formation energies, magnetic moments, and exchange coupling energies for the various defect structures considered herein (refer back to Fig. \ref{fig:structures}) are listed in Table \ref{tbl:calc}. In agreement with our spectroscopy measurements, the lowest formation energies (with respect to pure ZnO) are for isovalent single (1$S$) and antiferromagnetically coupled pairs (2$S$) of substitutional impurities, these formation energies are +0.66 eV and +0.82 eV, respectively. Combined substitutional and interstitial ($S$+$I$) impurities are not energetically favourable in the calculations (E$_{form}$ = 1.92 eV); this is notable since this arrangement can be the source of long-range ferromagnetism.~\cite{c20}

\begin{table}
\begin{tabular}{cccc}
 Defect Type & E$_{form}$ (eV) & M ($\mu_B$) & J (meV) \\
\hline
1$S$ & 0.66 & 3.90 & 0.0\\
2$S$ & 0.82 & 3.85 & -6.0\\
6$S$ & 1.40 & 3.78 & -2.0\\
\hline
1$S$ + $V_{O}$ & 5.59 & 3.91 & 0.0 \\
2$S$ + $V_{O}$ & 3.31 & 3.85 & -2.0 \\
\hline
1$S$ +$V_{Zn}$ & 6.09 & 0.00 (2+)& 0.0\\
2$S$ + $V_{Zn}$ & 2.09 & 4.06 & +1.0\\
6$S$ + $V_{Zn}$ & 1.04 & 3.92 (2+), 0.64 (3+) & +2$\sim$+4\\
\hline
1$S$ +$I_{O}$ & 0.06 & 3.29 & 0.0\\
2$S$ + $I_{O}$ & 0.17 & 3.91 & -1.0\\
\hline
1$S$ + $I_{Zn}$ & 3.60 & 3.54 & 0.0\\
2$S$ +$I_{Zn}$ & 2.47 & 3.56 & -1.0\\
Fe$_{surf}$ & -2.65 & 4.07 & 0.0\\
Fe$_{surf}$+O& -4.16 & 3.15 & 0.0\\
$I$ & 3.26 & 2.21 & 0.0\\
$S$+$I$ & 1.92 & 3.76($S$), 1.84($I$) & +2\\
2$S$ + 1$S$(Cu) & 0.56 & 3.42 & +7
\end{tabular}
\caption{\label{tbl:calc} Calculated formation energies (E$_{form}$) with respect to pure ZnO, magnetic moments (M), and exchange coupling energies (J) for several defect structures. Refer back to Fig. \ref{fig:structures} for the defect notations.}
\end{table}

To study the nature of experimentally detected Fe$^{3+}$ in thin films, calculations for various combinations of substitutional iron atoms, and various point defects have been performed. These calculations suggest that combinations of implanted iron with oxygen and zinc vacancies ($V_{O}$, $V_{Zn}$), and interstitial zinc atoms ($I_{Zn}$) all have large energy costs (above 2 eV), furthermore, these structures favour only Fe$^{2+}$ based defects with magnetic moments of around 4$\mu_B$. As an aside, the valencies are determined by the occupancy of the Fe orbitals in our calculations --- essentially the Mulliken charges~\cite{mulliken_55} --- and rounding to the nearest integer. The calculated occupancy for the entire 3\textit{d} shell is based on an orbital radius of 2.2~\AA, however for the radii between 1.8~\AA~and 3.0~\AA~the occupancy varies by only $\pm 0.3$ e. 

Only a single substitutional iron impurity combined with interstitial oxygen (1$S$+$I_{O}$) leads to Fe$^{3+}$, and this structure has a very low formation energy (0.06 eV). One of the reasons why this structure is energetically favourable is the change in local geometry of the Fe site from the roughly tetrahedral coordination of Zn to something closer to an octahedral coordination that is typically found in iron oxides (see Fig. \ref{fig:structures} (g)). This configuration requires additional oxygen, while typically ZnO samples have oxygen deficiencies, which are caused by the low formation energy of oxygen vacancies as compared to excess interstitial oxygen.~\cite{c21} The calculations indicate that in Fe-doped ZnO, the ratio of formation energies for interstitial oxygen ($I_{O}$) and oxygen vacancies ($V_{O}$) is just the opposite (0.06 eV for Fe-substitution and interstitial oxygen compared to the \mbox{5.59 eV} required for Fe-substitution and an oxygen vacancy, as shown in Table \ref{tbl:calc}). Therefore, the appearance of Fe$^{3+}$ can occur due to the favourable formation energies of interstitial oxygen. In addition to this, the calculations for iron substitution at the surface layer also have low formation energies (-2.65 eV for surface iron substitution, and \mbox{-4.16 eV} for surface iron substitution with oxygen pacification, as shown in Table \ref{tbl:calc}). This energetically favourable oxidation of surface iron impurities could be the origin of Fe$^{3+}$ in thin film samples, and this result is in agreement with our measurements and the measurements of others.~\cite{c23}

At this point we should make some comments on the ``bulk'' and ``surface'' aspects of these samples: As previously mentioned, our spectroscopic data quite plainly show a difference between the thin films and the pellets, however both of these samples are thicker than the ion implantation depth. Why then, should the surface of a Fe-doped ZnO thin film have significantly more Fe$^{3+}$ sites than the surface of a Fe-doped ZnO  pellet? The reason probably lies in the different surface geometries of these samples. The ZnO pellet has considerable surface roughness and no preferred orientation, while the thin film has less than 3 nm root-mean-square surface roughness and the polycrystalline grains strongly prefer to be oriented with the crystalline $c$-axis perpendicular to the substrate surface.~\cite{c8} Because of this, the average surface Fe substitution site in the thin film will quite closely resemble our DFT model structure [refer back to Fig.~\ref{fig:structures}(i)], while the average surface Fe substitution site in the pellet will have no preferred geometry. Indeed, because the polycrystalline grains in the pellet would have a much rougher surface than in the thin film, we may expect that significant surface reconstruction can occur local to the surface Fe substitution sites, leading to geometries quite similar to the energetically preferred Fe$^{2+}$ sites found in the bulk. In fact, it was previously found that preferential orientation of nanocrystalline grains of ZnO in thin films increases the chemical sensitivity of ZnO surfaces for vapor sensing.~\cite{wang_09}

\begin{figure}
\begin{center}
\includegraphics[width=3in]{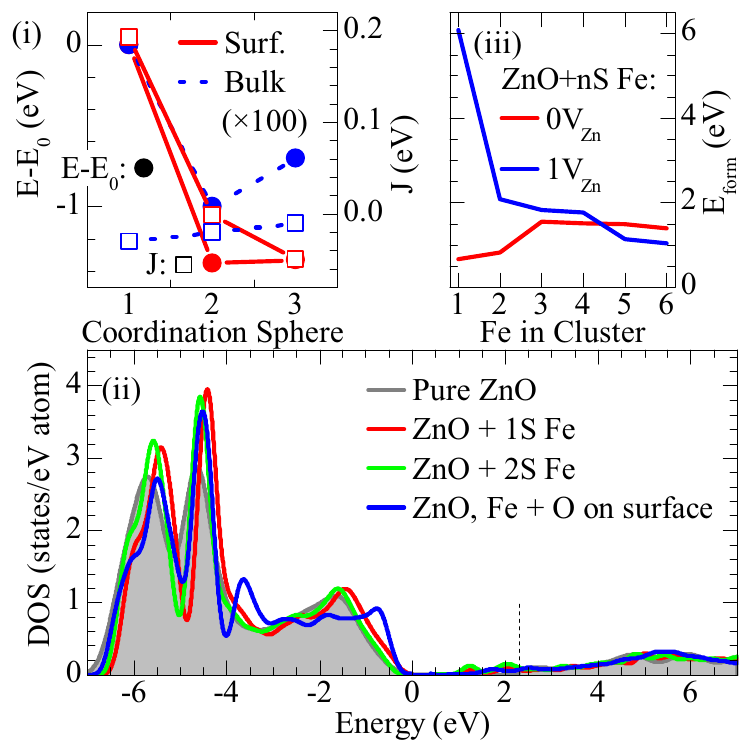}
\caption{Cluster calculations of ZnO with various defects: (i) The difference in the total energies and the exchange interactions between an iron substitution on the surface and another surface iron substitution (labelled ``surf''), or another iron substitution in the bulk (labelled ``bulk''), as a function of the distance between the iron impurities (expressed in terms coordination spheres). (ii) The DOS for several different structures, the dotted line indicates the bottom of the conduction band for pure ZnO (a calculated band gap of 2.3 eV). (iii) The formation energy as a function of the number of substitutional iron impurities with and without a zinc vacancy. (Color online.)}
\label{fig:calc_1}
\end{center}
\end{figure}

To analyze long-range super-exchange interactions for the single impurities in the presence of various point defects, the supercell was doubled along one of the axes and the exchange energies were calculated. These exchange energies were all extremely low (below 1 meV), and suggest the absence of long range magnetic interactions. It is therefore concluded that the iron implantation in the bulk induces 1$S$ and 2$S$ defect configurations; the former creates paramagnetic centres, while the latter creates antiferromagnetically coupled pairs, similar to codoped ZnO powders.~\cite{c13} The exchange interactions between the distance 1$S$ and 2$S$ impurities were always zero.

To investigate the possibility of substitutional iron atoms clustering in the vicinity of the ZnO surface, calculations on two possible types of impurity pairs were performed: the first is a two-substitutional impurity on a surface passivated by oxygen (Fe$^{3+}$ --- Fe$^{3+}$ exchange), while the second is a surface impurity passivated by oxygen and an Fe$^{2+}$ substitutional impurity in the bulk. As shown in Fig. \ref{fig:calc_1} (i), the first structure is quite energetically unfavourable when the second Fe$^{3+}$ impurity is in the first coordination sphere of the original impurity; this suggests that Fe$^{3+}$ impurities are distributed uniformly on the ZnO surface without any clustering. The exchange interactions for a second surface impurity in the second or third coordination sphere are antiferromagnetic (\mbox{J $<$ 0}, as shown in Fig. \ref{fig:calc_1} (i)); ferromagnetism can only be realized by the energetically unfavourable clustering of surface iron impurities. 

For the second structure, the formation energy is weakly dependent on the distance between the surface impurity and the bulk impurity, and the exchange energy is likewise weak and always antiferromagnetic  (see Fig. \ref{fig:calc_1}. Note that the formation energies and exchange interaction energies for the ``bulk'' plot have been amplified by a factor of 100). The calculations therefore suggest that Fe$^{3+}$ is uniformly distributed on the surface of ZnO (this is attractive for catalysis because the entire surface area is available), and ferromagnetism is not possible for isovalent substitution of Zn by Fe atoms. In fact, the latter point may help explain some recent experimental findings in which it is proposed that codoping with iron and copper is essential to the realization of ferromagnetism.~\cite{c24,c7,c26} While it has been shown that ferromagnetic ordering is possible for some structural defects (namely $S + I$ and 2$S + V_{Zn}$), we have determined that these structures are not energetically favourable. 

The calculated DOS, shown in Fig. \ref{fig:calc_1} (ii), suggest that Fe-doping in the 1$S$ and 2$S$ structures reduces the band gap to 0.47 eV and 0.52 eV, respectively. This is only about 20\% of the calculated gap for pure ZnO (2.3 eV). It should be pointed out that while DFT calculations are known to significantly underestimate the magnitude of the band gap,~\cite{dufek_94} we expect that the qualitative trend of band gap reduction is accurate. The calculated DOS shows that hybridization between the O 2$p$-Fe 3$d$ states leads to a small shift in the oxygen 2$p$ band (this is also suggested in the O $K$-edge XES measurements, refer back for Fig. \ref{fig:o_sxs}), this hybridization also changes the position of the Zn 3$d$ band, providing the experimentally detected smearing of the corresponding peaks in the XPS valence band spectra (see Fig. \ref{fig:xps}) and the O $2p$ --- Zn $3d$-hybridization peak in the O $K$-edge XES spectra (see the feature near 520 eV in Fig. \ref{fig:o_sxs}).

The appearance of ferromagnetism in Fe-doped ZnO has previously been attributed to a secondary magnetic phase of ZnFe$_2$O$_4$ (containing Fe$^{3+}$).~\cite{c27} To examine this, the formation energies for clusters with $n$ substitutional iron impurities with and without Zn vacancies were calculated($nS$, and $nS+V_{Zn}$). These formation energies, shown in Fig. \ref{fig:calc_1}(iii), indicate that for large clusters of Fe (note that this is Fe clustering in the bulk, not on the surface), Zn vacancies become increasingly energetically favourable. This is, for a cluster with 6 iron atoms the formation energy decreases from 1.40 eV to 1.04 eV per defect with a Zn vacancy (i.e. from 6$S$ to 6$S+V_{Zn}$). This makes the formation of a secondary ZnFe$_2$O$_4$ phase very probable in the iron-rich areas of doped ZnO. The magnetic structure of a 6$S+V_{Zn}$ cluster suggests that two iron atoms convert to a Fe$^{3+}$ low spin configuration, and the exchange interactions within this cluster are positive and induce ferromagnetism (see Table \ref{tbl:calc}). However, for long distances (about 1 nm) the exchange interaction between clusters remains negligible (less than 1 meV), in this way only the formation of larger sizes of similar clusters can be a source of ferromagnetism in Fe-doped ZnO.

\begin{figure}
\begin{center}
\includegraphics[width=3in]{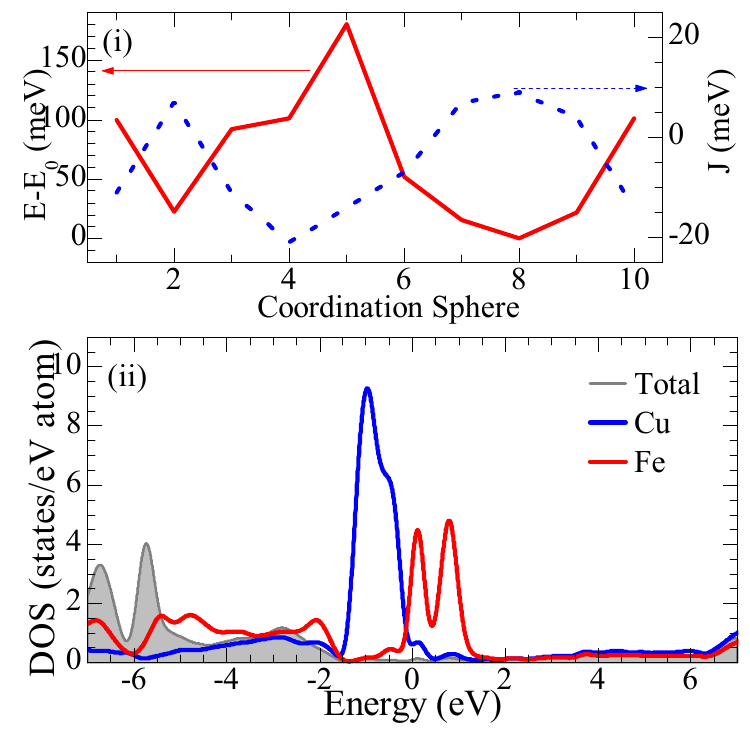}
\caption{Cluster calculations of ZnO with iron and copper impurities: (i) The energy difference between the current and most energetically favourable structure and the exchange interaction as a function of the distance between iron impurities (expressed in terms of coordination spheres). (ii) The DOS for the most energetically favourable configuration of ZnO:Fe,Cu. (Color online.)}
\label{fig:calc_2}
\end{center}
\end{figure}

One other possibility for the realization of ferromagnetism in Fe-doped ZnO discussed in the literature is the case of copper codoping.~\cite{c24} To model the effect of copper codoping, single copper substitution was performed and the distance between two substituted iron defects was varied. In this situation there was a weak variation (about 0.1 eV) in the formation energy of the system as a function of the change in distance between iron impurities, as shown in Fig. \ref{fig:calc_2}(i). However there is a substantial change in the magnetic properties, and the antiferromagnetic super-exchange interaction becomes similar to a Ruderman-Kittel-Kasuya-Yosida (RKKY) interaction (see Table \ref{tbl:calc}, and Fig. \ref{fig:calc_2} (ii)).~\cite{yafet_87} The  RKKY interaction couples magnetic moments indirectly through conduction electrons and is the dominant exchange interaction in metals that have little overlap between magnetic electrons. The DOS of the most energetically favourable Fe Cu-codoped ZnO system is also significantly different from that of Fe-doped ZnO, as shown in Fig. \ref{fig:calc_2} (ii). In particular, this occurs when the system is metallic and has very localized unoccupied Fe 3$d$ orbitals immediately above the Fermi level (compare to Fig. \ref{fig:calc_1}(ii)). These changes occur because Cu$^{1+}$ has a 3$d^{10}$4$s^0$ configuration (similar to Zn$^{2+}$). This allows the formation of a single hole that should convert one of the iron atoms to a 3+ (from the original 2+) oxidation state. However, the iron defects are equivalent, so the hole is delocalized between the two, shifting the conduction band (see Fig. \ref{fig:calc_2}(ii) and reducing the magnetic moment of the substitutional iron impurities (see Table \ref{tbl:calc}). This effectively creates charge carriers in the system, creating an RKKY-type exchange interaction. The calculations suggest that this is the most energetically favourable configuration of iron impurities, and it is estimated to have a Curie temperature of 463 K using the Ising model ($T_\mathrm{C}$ = $\frac{3n}{k_B} J(J+1)$ where $J$ is the exchange energy, $n$ is the number of exchanges, and $k_B$ is the usual Boltzmann constant). This Curie temperature, and the general hole-type conductivity of the system, are in good agreement with experimental results.~\cite{c24}

\section{Conclusions}

To summarize, we have studied the local structure of Fe-impurity atoms in pellet and thin film ZnO samples with various x-ray spectroscopic techniques and DFT calculations. It is found that isovalent (and isostructural) substitution (Fe$^{2+} \rightarrow$ Zn$^{2+}$) takes place in both types of materials. In thin films, additional heterovalent Fe$^{3+}$ substitutions (again for Zn) were found primarily near the surface, and are likely due to the relatively pristine crystalline order found at the surfaces of these films. Our calculations indicate that the formation energies of single and double substitutional impurities are low, leading to the formation of paramagnetic centers or antiferromagnetic ordering, respectively. These findings assist in explaining why experimental observations of ferromagnetism are seen in Fe-doped bulk ZnO only after additional Cu-doping; the reason for this is because ferromagnetic ordering occurs only in energetically unfavourable iron defect geometries.

In addition, we find that the presence of Fe$^{3+}$ ions can be attributed to the formation of a secondary magnetic phase of ZnFe$_2$O$_4$, the presence of interstitial oxygen near the iron substitution site, or additional oxidation of iron defects on the surface of ZnO. For the purpose of developing spintronic devices, Fe-doped ZnO is not a promising material because reproducible experimental results can only be achieved in bulk samples where ferromagnetism seems dependent on Cu-codoping. On the other hand, Fe-doped ZnO may be a promising photocatalytic substrate because Fe-doping reduces the band gap; and in this regime the high level of doping used in the samples herein is appropriate.
\acknowledgement

We gratefully acknowledge support from the Natural Sciences and Engineering Research Council of Canada (NSERC) and the Canada Research Chair program. This work was done with partial support of Ural Division of Russian Academy of Sciences (Project 
12-I-2-2040) and the Russian Foundation for Basic Research (Project 13-08-00059). The Advanced Light Source is supported by the Director, Office of Science, Office of Basic Energy Sciences, of the U. S. Department of Energy under Contract No. DE-AC02-05CH11231. The Canadian Light Source is supported by NSERC, the National Research Council (NSC) Canada, the Canadian Institute of Health Research (CIHR), the Province of Saskatchewan, Western Economic Diversification Canada, and the University of Saskatchewan. DWB acknowledges the computation support from CAC of KIAS.

\bibliography{ZnO_Fe_film_JPCC}

\providecommand{\latin}[1]{#1}
\providecommand*\mcitethebibliography{\thebibliography}
\csname @ifundefined\endcsname{endmcitethebibliography}
  {\let\endmcitethebibliography\endthebibliography}{}
\begin{mcitethebibliography}{53}
\providecommand*\natexlab[1]{#1}
\providecommand*\mciteSetBstSublistMode[1]{}
\providecommand*\mciteSetBstMaxWidthForm[2]{}
\providecommand*\mciteBstWouldAddEndPuncttrue
  {\def\EndOfBibitem{\unskip.}}
\providecommand*\mciteBstWouldAddEndPunctfalse
  {\let\EndOfBibitem\relax}
\providecommand*\mciteSetBstMidEndSepPunct[3]{}
\providecommand*\mciteSetBstSublistLabelBeginEnd[3]{}
\providecommand*\EndOfBibitem{}
\mciteSetBstSublistMode{f}
\mciteSetBstMaxWidthForm{subitem}{(\alph{mcitesubitemcount})}
\mciteSetBstSublistLabelBeginEnd
  {\mcitemaxwidthsubitemform\space}
  {\relax}
  {\relax}

\bibitem[Pearton \latin{et~al.}(2003)Pearton, Abernathy, Norton, Hebard, Park,
  Boatner, and Budai]{c1}
Pearton,~S.; Abernathy,~C.; Norton,~D.; Hebard,~A.; Park,~Y.; Boatner,~L.;
  Budai,~J. {Advances in Wide Bandgap Materials for Semiconductor Spintronics}.
  \emph{Mater. Sci. Eng., R} \textbf{2003}, \emph{40}, 137--168\relax
\mciteBstWouldAddEndPuncttrue
\mciteSetBstMidEndSepPunct{\mcitedefaultmidpunct}
{\mcitedefaultendpunct}{\mcitedefaultseppunct}\relax
\EndOfBibitem
\bibitem[Xu \latin{et~al.}(2008)Xu, Yang, Liu, Huang, Lee, Cho, and
  Wang]{xu_08}
Xu,~C.; Yang,~K.; Liu,~Y.; Huang,~L.; Lee,~H.; Cho,~J.; Wang,~H. {Buckling and
  Ferromagnetism of Aligned Cr-Doped ZnO Nanorods}. \emph{J. Phys. Chem. C}
  \textbf{2008}, \emph{112}, 19236--19241\relax
\mciteBstWouldAddEndPuncttrue
\mciteSetBstMidEndSepPunct{\mcitedefaultmidpunct}
{\mcitedefaultendpunct}{\mcitedefaultseppunct}\relax
\EndOfBibitem
\bibitem[Ekambaram(2008)]{c2}
Ekambaram,~S. {Photoproduction of Clean H2 or O2 from Water using Oxide
  Semiconductors in Presence of Sacrificial Reagent}. \emph{J. Alloys and
  Compd.} \textbf{2008}, \emph{448}, 238--245\relax
\mciteBstWouldAddEndPuncttrue
\mciteSetBstMidEndSepPunct{\mcitedefaultmidpunct}
{\mcitedefaultendpunct}{\mcitedefaultseppunct}\relax
\EndOfBibitem
\bibitem[Fujishima and Honda(1972)Fujishima, and Honda]{c3}
Fujishima,~A.; Honda,~K. {Electrochemical Photolysis of Water at a
  Semiconductor Electrode}. \emph{Nature} \textbf{1972}, \emph{238},
  37--38\relax
\mciteBstWouldAddEndPuncttrue
\mciteSetBstMidEndSepPunct{\mcitedefaultmidpunct}
{\mcitedefaultendpunct}{\mcitedefaultseppunct}\relax
\EndOfBibitem
\bibitem[Catlow \latin{et~al.}(2010)Catlow, Guo, Miskufova, Shevlin, Smith,
  Sokol, Walsh, Wilson, and Woodley]{c4}
Catlow,~C. R.~A.; Guo,~Z.~X.; Miskufova,~M.; Shevlin,~S.~A.; Smith,~A. G.~H.;
  Sokol,~A.~A.; Walsh,~A.; Wilson,~D.~J.; Woodley,~S.~M. {Advances in
  Computational Studies of Energy Materials.} \emph{Philos. Trans. R. Soc.
  London, Ser. A} \textbf{2010}, \emph{368}, 3379--456\relax
\mciteBstWouldAddEndPuncttrue
\mciteSetBstMidEndSepPunct{\mcitedefaultmidpunct}
{\mcitedefaultendpunct}{\mcitedefaultseppunct}\relax
\EndOfBibitem
\bibitem[Lee \latin{et~al.}(2010)Lee, Han, Jung, Shin, Lee, Noh, Lee, Cho, Lee,
  Kim, and {et al.}]{lee_10}
Lee,~S.-H.; Han,~S.-H.; Jung,~H.~S.; Shin,~H.; Lee,~J.; Noh,~J.-H.; Lee,~S.;
  Cho,~I.-S.; Lee,~J.-K.; Kim,~J.; {et al.}, {Al-Doped ZnO Thin Film: A New
  Transparent Conducting Layer for ZnO Nanowire-Based Dye-Sensitized Solar
  Cells}. \emph{J. Phys. Chem. C} \textbf{2010}, \emph{114}, 7185--7189\relax
\mciteBstWouldAddEndPuncttrue
\mciteSetBstMidEndSepPunct{\mcitedefaultmidpunct}
{\mcitedefaultendpunct}{\mcitedefaultseppunct}\relax
\EndOfBibitem
\bibitem[Raj \latin{et~al.}(2013)Raj, Prabakar, Karthick, Hemalatha, Son, and
  Kim]{raj_13}
Raj,~C.~J.; Prabakar,~K.; Karthick,~S.~N.; Hemalatha,~K.~V.; Son,~M.-K.;
  Kim,~H.-J. {Banyan Root Structured Mg-Doped ZnO Photoanode Dye-Sensitized
  Solar Cells}. \emph{J. Phys. Chem. C} \textbf{2013}, \emph{117},
  2600--2607\relax
\mciteBstWouldAddEndPuncttrue
\mciteSetBstMidEndSepPunct{\mcitedefaultmidpunct}
{\mcitedefaultendpunct}{\mcitedefaultseppunct}\relax
\EndOfBibitem
\bibitem[Kikoin(2009)]{c5}
Kikoin,~K. {Ferromagnetic Ordering in Dilute Magnetic Dielectrics with and
  without Free Carriers}. \emph{Journal of Magn. Magn. Mater.} \textbf{2009},
  \emph{321}, 702--705\relax
\mciteBstWouldAddEndPuncttrue
\mciteSetBstMidEndSepPunct{\mcitedefaultmidpunct}
{\mcitedefaultendpunct}{\mcitedefaultseppunct}\relax
\EndOfBibitem
\bibitem[Hong \latin{et~al.}(2005)Hong, Sakai, Huong, Poirot, and Ruyter]{c6}
Hong,~N.~H.; Sakai,~J.; Huong,~N.~T.; Poirot,~N.; Ruyter,~A. {Role of Defects
  in Tuning Ferromagnetism in Diluted Magnetic Oxide Thin Films}. \emph{Phys.
  Rev. B} \textbf{2005}, \emph{72}, 045336--5\relax
\mciteBstWouldAddEndPuncttrue
\mciteSetBstMidEndSepPunct{\mcitedefaultmidpunct}
{\mcitedefaultendpunct}{\mcitedefaultseppunct}\relax
\EndOfBibitem
\bibitem[Sato and Katayama-Yoshida(2000)Sato, and Katayama-Yoshida]{sato_00}
Sato,~K.; Katayama-Yoshida,~H. {Material Design for Transparent Ferromagnets
  with ZnO-Based Magnetic Semiconductors}. \emph{Japn. J. Appl. Phys.}
  \textbf{2000}, \emph{39}, L555--L558\relax
\mciteBstWouldAddEndPuncttrue
\mciteSetBstMidEndSepPunct{\mcitedefaultmidpunct}
{\mcitedefaultendpunct}{\mcitedefaultseppunct}\relax
\EndOfBibitem
\bibitem[Jin \latin{et~al.}(2001)Jin, Fukumura, Kawasaki, Ando, Saito,
  Sekiguchi, Yoo, Murakami, Matsumoto, Hasegawa, and {et al.}]{jin_01}
Jin,~Z.; Fukumura,~T.; Kawasaki,~M.; Ando,~K.; Saito,~H.; Sekiguchi,~T.;
  Yoo,~Y.~Z.; Murakami,~M.; Matsumoto,~Y.; Hasegawa,~T.; {et al.}, {High
  Throughput Fabrication of Transition-Metal-Doped Epitaxial ZnO Thin Films: A
  Series of Oxide-Diluted Magnetic Semiconductors and their Properties}.
  \emph{Appl. Phys. Lett.} \textbf{2001}, \emph{78}, 3824--3826\relax
\mciteBstWouldAddEndPuncttrue
\mciteSetBstMidEndSepPunct{\mcitedefaultmidpunct}
{\mcitedefaultendpunct}{\mcitedefaultseppunct}\relax
\EndOfBibitem
\bibitem[Inamdar \latin{et~al.}(2011)Inamdar, Pathak, Dubenko, Ali, and
  Mahamuni]{inamdar_11}
Inamdar,~D.~Y.; Pathak,~A.~K.; Dubenko,~I.; Ali,~N.; Mahamuni,~S. {Room
  Temperature Ferromagnetism and Photoluminescence of Fe Doped ZnO
  Nanocrystals}. \emph{J. Phys. Chem. C} \textbf{2011}, \emph{115},
  23671--23676\relax
\mciteBstWouldAddEndPuncttrue
\mciteSetBstMidEndSepPunct{\mcitedefaultmidpunct}
{\mcitedefaultendpunct}{\mcitedefaultseppunct}\relax
\EndOfBibitem
\bibitem[Hong \latin{et~al.}(2007)Hong, Sakai, and Briz\'{e}]{hong_07}
Hong,~N.~H.; Sakai,~J.; Briz\'{e},~V. {Observation of Ferromagnetism at Room
  Temperature in ZnO Thin Films}. \emph{J. Phys.: Condens. Matt.}
  \textbf{2007}, \emph{19}, 036219--6\relax
\mciteBstWouldAddEndPuncttrue
\mciteSetBstMidEndSepPunct{\mcitedefaultmidpunct}
{\mcitedefaultendpunct}{\mcitedefaultseppunct}\relax
\EndOfBibitem
\bibitem[Karmakar \latin{et~al.}(2007)Karmakar, Mandal, Kadam, Paulose,
  Rajarajan, Nath, Das, Dasgupta, and Das]{karmakar_07}
Karmakar,~D.; Mandal,~S.~K.; Kadam,~R.~M.; Paulose,~P.~L.; Rajarajan,~A.~K.;
  Nath,~T.~K.; Das,~A.~K.; Dasgupta,~I.; Das,~G.~P. {Ferromagnetism in Fe-doped
  ZnO Nanocrystals: Experiment and Theory}. \emph{Phys. Rev. B} \textbf{2007},
  \emph{75}, 144404--14\relax
\mciteBstWouldAddEndPuncttrue
\mciteSetBstMidEndSepPunct{\mcitedefaultmidpunct}
{\mcitedefaultendpunct}{\mcitedefaultseppunct}\relax
\EndOfBibitem
\bibitem[Kohan \latin{et~al.}(2000)Kohan, Ceder, Morgan, and {de Walle}]{c21}
Kohan,~A.~F.; Ceder,~G.; Morgan,~D.; {de Walle},~C. G.~V. {First-Principles
  Study of Native Point Defects in ZnO}. \emph{Phys. Rev. B} \textbf{2000},
  \emph{61}, 15019--15027\relax
\mciteBstWouldAddEndPuncttrue
\mciteSetBstMidEndSepPunct{\mcitedefaultmidpunct}
{\mcitedefaultendpunct}{\mcitedefaultseppunct}\relax
\EndOfBibitem
\bibitem[Jeong \latin{et~al.}(2004)Jeong, Han, Park, and Lee]{c7}
Jeong,~Y.; Han,~S.; Park,~J.; Lee,~Y. {A Critical Examination of Room
  Temperature Ferromagnetism in Transition Metal-Doped Oxide Semiconductors}.
  \emph{J. Magn. Magn. Mater.} \textbf{2004}, \emph{272-276}, 1976--1980\relax
\mciteBstWouldAddEndPuncttrue
\mciteSetBstMidEndSepPunct{\mcitedefaultmidpunct}
{\mcitedefaultendpunct}{\mcitedefaultseppunct}\relax
\EndOfBibitem
\bibitem[de~Groot(2005)]{degroot_05}
de~Groot,~F. {Multiplet Effects in X-ray Spectroscopy}. \emph{Coordin. Chem.
  Rev.} \textbf{2005}, \emph{249}, 31--63\relax
\mciteBstWouldAddEndPuncttrue
\mciteSetBstMidEndSepPunct{\mcitedefaultmidpunct}
{\mcitedefaultendpunct}{\mcitedefaultseppunct}\relax
\EndOfBibitem
\bibitem[Cui \latin{et~al.}(2012)Cui, Zhang, Wang, Yang, Kuang, Sun, and
  Han]{c8}
Cui,~L.; Zhang,~H.~Y.; Wang,~G.~G.; Yang,~F.~X.; Kuang,~X.~P.; Sun,~R.;
  Han,~J.~C. {Effect of Annealing Temperature and Annealing Atmosphere on the
  Structure and Optical Properties of ZnO Thin Films on Sapphire (0001)
  Substrates by Magnetron Sputtering}. \emph{Appl. Surf. Sci.} \textbf{2012},
  \emph{258}, 2479--2485\relax
\mciteBstWouldAddEndPuncttrue
\mciteSetBstMidEndSepPunct{\mcitedefaultmidpunct}
{\mcitedefaultendpunct}{\mcitedefaultseppunct}\relax
\EndOfBibitem
\bibitem[Ziegler \latin{et~al.}(2008)Ziegler, Biersack, and Ziegler]{srim_ref}
Ziegler,~J.~F.; Biersack,~J.~P.; Ziegler,~M.~D. \emph{{SRIM The Stopping and
  Range of Ions in Matter}}; www.srim.org, 2008\relax
\mciteBstWouldAddEndPuncttrue
\mciteSetBstMidEndSepPunct{\mcitedefaultmidpunct}
{\mcitedefaultendpunct}{\mcitedefaultseppunct}\relax
\EndOfBibitem
\bibitem[Tougaard(1987)]{c9}
Tougaard,~S. {Low Energy Inelastic Electron Scattering Properties of Noble and
  Transition Metals}. \emph{Solid State Commun.} \textbf{1987}, \emph{61},
  547--549\relax
\mciteBstWouldAddEndPuncttrue
\mciteSetBstMidEndSepPunct{\mcitedefaultmidpunct}
{\mcitedefaultendpunct}{\mcitedefaultseppunct}\relax
\EndOfBibitem
\bibitem[Moulder \latin{et~al.}(1992)Moulder, Stickle, Sobol, and Bomben]{c10}
Moulder,~J.; Stickle,~W.; Sobol,~P.; Bomben,~K. In \emph{Handbook of X-ray
  Photoelectron Spectroscopy}; Chastain,~J., Ed.; Perkin-Elmer Corporation,
  1992; p 261\relax
\mciteBstWouldAddEndPuncttrue
\mciteSetBstMidEndSepPunct{\mcitedefaultmidpunct}
{\mcitedefaultendpunct}{\mcitedefaultseppunct}\relax
\EndOfBibitem
\bibitem[Jia \latin{et~al.}(1995)Jia, Callcott, Yurkas, Ellis, and
  Himpsel]{jia_95}
Jia,~J.~J.; Callcott,~T.~A.; Yurkas,~J.; Ellis,~A.~W.; Himpsel,~F.~J. {First
  Experimental Results from EM / TENN / TULANE / LLNL / LBL Undulator Beamline
  at the Advanced Light Source}. \emph{Rev. Sci. Instr.} \textbf{1995},
  \emph{66}, 1394--1397\relax
\mciteBstWouldAddEndPuncttrue
\mciteSetBstMidEndSepPunct{\mcitedefaultmidpunct}
{\mcitedefaultendpunct}{\mcitedefaultseppunct}\relax
\EndOfBibitem
\bibitem[Regier \latin{et~al.}(2007)Regier, Krochak, Sham, Hu, Thompson, and
  Blyth]{regier_07}
Regier,~T.; Krochak,~J.; Sham,~T.~K.; Hu,~Y.~F.; Thompson,~J.; Blyth,~R. I.~R.
  \emph{Nucl. Instrum. Meth. A} \textbf{2007}, \emph{582}, 93--95\relax
\mciteBstWouldAddEndPuncttrue
\mciteSetBstMidEndSepPunct{\mcitedefaultmidpunct}
{\mcitedefaultendpunct}{\mcitedefaultseppunct}\relax
\EndOfBibitem
\bibitem[Ordej{\'{o}}n \latin{et~al.}(1996)Ordej{\'{o}}n, Artacho, and
  Soler]{SIESTA_1}
Ordej{\'{o}}n,~P.; Artacho,~E.; Soler,~J.~M. {Self-Consistent Order-$N$
  Density-Functional Calculations for Very Large Systems}. \emph{Phys. Rev. B}
  \textbf{1996}, \emph{53}, R10441--R10444\relax
\mciteBstWouldAddEndPuncttrue
\mciteSetBstMidEndSepPunct{\mcitedefaultmidpunct}
{\mcitedefaultendpunct}{\mcitedefaultseppunct}\relax
\EndOfBibitem
\bibitem[Soler \latin{et~al.}(2002)Soler, Artacho, J, Garc{\'{i}}a, Junquera,
  Ordej{\'{o}}n, and S{\'{a}}nchez-Portal]{SIESTA_2}
Soler,~J.~M.; Artacho,~E.; J,~D.~G.; Garc{\'{i}}a,~A.; Junquera,~J.;
  Ordej{\'{o}}n,~P.; S{\'{a}}nchez-Portal,~D. {The SIESTA Method for Ab Initio
  Order-N Materials Simulation}. \emph{J. Phys.: Condens. Matter}
  \textbf{2002}, \emph{14}, 2745--2779\relax
\mciteBstWouldAddEndPuncttrue
\mciteSetBstMidEndSepPunct{\mcitedefaultmidpunct}
{\mcitedefaultendpunct}{\mcitedefaultseppunct}\relax
\EndOfBibitem
\bibitem[Chang \latin{et~al.}(2007)Chang, Kurmaev, Boukhvalov, Finkelstein,
  Colis, Pedersen, Moewes, and Dinia]{c13}
Chang,~G.~S.; Kurmaev,~E.~Z.; Boukhvalov,~D.~W.; Finkelstein,~L.~D.; Colis,~S.;
  Pedersen,~T.~M.; Moewes,~A.; Dinia,~A. {Effect of Co and O Defects on the
  Magnetism in Co-doped ZnO: Experiment and Theory}. \emph{Phys. Rev. B}
  \textbf{2007}, \emph{75}, 195215--7\relax
\mciteBstWouldAddEndPuncttrue
\mciteSetBstMidEndSepPunct{\mcitedefaultmidpunct}
{\mcitedefaultendpunct}{\mcitedefaultseppunct}\relax
\EndOfBibitem
\bibitem[Perdew \latin{et~al.}(1996)Perdew, Burke, and Ernzerhof]{c14a}
Perdew,~J.~P.; Burke,~K.; Ernzerhof,~M. {Generalized Gradient Approximation
  Made Simple.} \emph{Phys. Rev. Lett.} \textbf{1996}, \emph{77},
  3865--3868\relax
\mciteBstWouldAddEndPuncttrue
\mciteSetBstMidEndSepPunct{\mcitedefaultmidpunct}
{\mcitedefaultendpunct}{\mcitedefaultseppunct}\relax
\EndOfBibitem
\bibitem[Monkhorst and Pack(1976)Monkhorst, and Pack]{c15a}
Monkhorst,~H.~J.; Pack,~J.~D. {Special Points for Brillouin-Zone Integrations}.
  \emph{Phys. Rev. B} \textbf{1976}, \emph{13}, 5188--5192\relax
\mciteBstWouldAddEndPuncttrue
\mciteSetBstMidEndSepPunct{\mcitedefaultmidpunct}
{\mcitedefaultendpunct}{\mcitedefaultseppunct}\relax
\EndOfBibitem
\bibitem[Garc{\'{i}}a-Mota \latin{et~al.}(2012)Garc{\'{i}}a-Mota, Vojvodic,
  Abild-Pedersen, and N{\o}rskov]{garcia_12}
Garc{\'{i}}a-Mota,~M.; Vojvodic,~A.; Abild-Pedersen,~F.; N{\o}rskov,~J.~K.
  {Electronic Origin of the Surface Reactivity of Transition-Metal-Doped TiO2
  (110)}. \emph{J. Phys. Chem. C} \textbf{2012}, \emph{117}, 460--465\relax
\mciteBstWouldAddEndPuncttrue
\mciteSetBstMidEndSepPunct{\mcitedefaultmidpunct}
{\mcitedefaultendpunct}{\mcitedefaultseppunct}\relax
\EndOfBibitem
\bibitem[Sato \latin{et~al.}(2010)Sato, Bergqvist, Kudrnovsk\'{y}, Dederichs,
  Eriksson, Turek, Sanyal, Bouzerar, Katayama-Yoshida, Dinh, and {et
  al.}]{sato_10}
Sato,~K.; Bergqvist,~L.; Kudrnovsk\'{y},~J.; Dederichs,~P.~H.; Eriksson,~O.;
  Turek,~I.; Sanyal,~B.; Bouzerar,~G.; Katayama-Yoshida,~H.; Dinh,~V.~A.; {et
  al.}, {First-Principles Theory of Dilute Magnetic Semiconductors}. \emph{Rev.
  Mod. Phys.} \textbf{2010}, \emph{82}, 1633--1690\relax
\mciteBstWouldAddEndPuncttrue
\mciteSetBstMidEndSepPunct{\mcitedefaultmidpunct}
{\mcitedefaultendpunct}{\mcitedefaultseppunct}\relax
\EndOfBibitem
\bibitem[Lany and Zunger(2008)Lany, and Zunger]{lany_08}
Lany,~S.; Zunger,~A. {Assessment of Correction Methods for the Band-gap Problem
  and for Finite-Size Effects in Supercell Defect Calculations: Case Studies
  for ZnO and GaAs}. \emph{Phys. Rev. B} \textbf{2008}, \emph{78},
  235104--25\relax
\mciteBstWouldAddEndPuncttrue
\mciteSetBstMidEndSepPunct{\mcitedefaultmidpunct}
{\mcitedefaultendpunct}{\mcitedefaultseppunct}\relax
\EndOfBibitem
\bibitem[Lany and Zunger(2010)Lany, and Zunger]{lany_10}
Lany,~S.; Zunger,~A. {Many-Body GW Calculation of the Oxygen Vacancy in ZnO}.
  \emph{Phys. Rev. B} \textbf{2010}, \emph{81}, 113201--4\relax
\mciteBstWouldAddEndPuncttrue
\mciteSetBstMidEndSepPunct{\mcitedefaultmidpunct}
{\mcitedefaultendpunct}{\mcitedefaultseppunct}\relax
\EndOfBibitem
\bibitem[Dufek \latin{et~al.}(1994)Dufek, Blaha, and Schwarz]{dufek_94}
Dufek,~P.; Blaha,~P.; Schwarz,~K. {Applications of Engel and Vosko's
  Generalized Gradient Approximation in Solids}. \emph{Phys. Rev. B}
  \textbf{1994}, \emph{50}, 7279--7283\relax
\mciteBstWouldAddEndPuncttrue
\mciteSetBstMidEndSepPunct{\mcitedefaultmidpunct}
{\mcitedefaultendpunct}{\mcitedefaultseppunct}\relax
\EndOfBibitem
\bibitem[Fujii \latin{et~al.}(1999)Fujii, {de Groot}, Sawatzky, Voogt, Hibma,
  and Okada]{c15b}
Fujii,~T.; {de Groot},~F. M.~F.; Sawatzky,~G.~A.; Voogt,~F.~C.; Hibma,~T.;
  Okada,~K. {In Situ XPS Analysis of various Iron Oxide Films Grown by
  NO2-Assisted Molecular-Beam Epitaxy}. \emph{Phys. Rev. B} \textbf{1999},
  \emph{59}, 3195--3202\relax
\mciteBstWouldAddEndPuncttrue
\mciteSetBstMidEndSepPunct{\mcitedefaultmidpunct}
{\mcitedefaultendpunct}{\mcitedefaultseppunct}\relax
\EndOfBibitem
\bibitem[{van der Laan} \latin{et~al.}(1988){van der Laan}, Thole, Sawatzky,
  and Verdaguer]{laan_88}
{van der Laan},~G.; Thole,~B.~T.; Sawatzky,~G.~A.; Verdaguer,~M. {Multiplet
  Structure in the ${L}_{2,3}$ X-ray-Absorption Spectra: A Fingerprint for
  High- and Low-Spin Ni$^{2+}$ Compounds}. \emph{Phys. Rev. B} \textbf{1988},
  \emph{37}, 6587--6589\relax
\mciteBstWouldAddEndPuncttrue
\mciteSetBstMidEndSepPunct{\mcitedefaultmidpunct}
{\mcitedefaultendpunct}{\mcitedefaultseppunct}\relax
\EndOfBibitem
\bibitem[Lee and Oh(1991)Lee, and Oh]{lee_91}
Lee,~G.; Oh,~S.-J. {Electronic Structures of NiO, CoO, and FeO Studied by 2p
  Core-Level X-ray Photoelectron Spectroscopy}. \emph{Phys. Rev. B}
  \textbf{1991}, \emph{43}, 14674--14682\relax
\mciteBstWouldAddEndPuncttrue
\mciteSetBstMidEndSepPunct{\mcitedefaultmidpunct}
{\mcitedefaultendpunct}{\mcitedefaultseppunct}\relax
\EndOfBibitem
\bibitem[Kotani and Shin(2001)Kotani, and Shin]{kotani_01}
Kotani,~A.; Shin,~S. {Resonant Inelastic X-ray Scattering Spectra for Electrons
  in Solids}. \emph{Rev. Mod. Phys.} \textbf{2001}, \emph{73}, 203--246\relax
\mciteBstWouldAddEndPuncttrue
\mciteSetBstMidEndSepPunct{\mcitedefaultmidpunct}
{\mcitedefaultendpunct}{\mcitedefaultseppunct}\relax
\EndOfBibitem
\bibitem[Bocquet \latin{et~al.}(1992)Bocquet, Mizokawa, Saitoh, Namatame, and
  Fujimori]{bocquet_92}
Bocquet,~A.; Mizokawa,~T.; Saitoh,~T.; Namatame,~H.; Fujimori,~A. {Electronic
  Structure of 3d-Transition-Metal Compounds by Analysis of the 2p Core-Level
  Photoemission Spectra}. \emph{Physical Review B} \textbf{1992}, \emph{46},
  3771--3784\relax
\mciteBstWouldAddEndPuncttrue
\mciteSetBstMidEndSepPunct{\mcitedefaultmidpunct}
{\mcitedefaultendpunct}{\mcitedefaultseppunct}\relax
\EndOfBibitem
\bibitem[Brundle \latin{et~al.}(1977)Brundle, Chuang, and Wandelt]{brundle_77}
Brundle,~C.; Chuang,~T.; Wandelt,~K. {Core and Valence Level Photoemission
  Studies of Iron Oxide Surfaces and the Oxidation of Iron}. \emph{Surface
  Science} \textbf{1977}, \emph{68}, 459--468\relax
\mciteBstWouldAddEndPuncttrue
\mciteSetBstMidEndSepPunct{\mcitedefaultmidpunct}
{\mcitedefaultendpunct}{\mcitedefaultseppunct}\relax
\EndOfBibitem
\bibitem[Yamashita and Hayes(2008)Yamashita, and Hayes]{c14b}
Yamashita,~T.; Hayes,~P. {Analysis of XPS Spectra of Fe2+ and Fe3+ Ions in
  Oxide Materials}. \emph{Appl. Surf. Sci.} \textbf{2008}, \emph{254},
  2441--2449\relax
\mciteBstWouldAddEndPuncttrue
\mciteSetBstMidEndSepPunct{\mcitedefaultmidpunct}
{\mcitedefaultendpunct}{\mcitedefaultseppunct}\relax
\EndOfBibitem
\bibitem[Skorikov \latin{et~al.}(2013)Skorikov, Korotin, Kurmaev, and
  Cholakh]{c17}
Skorikov,~N.~A.; Korotin,~M.~A.; Kurmaev,~E.~Z.; Cholakh,~S.~O. {Computer
  Simulation of the Energy Gap in ZnO- and TiO2-Based Semiconductor
  Photocatalysts}. \emph{J. Exp. Theor. Phys.} \textbf{2013}, \emph{115},
  1048--1054\relax
\mciteBstWouldAddEndPuncttrue
\mciteSetBstMidEndSepPunct{\mcitedefaultmidpunct}
{\mcitedefaultendpunct}{\mcitedefaultseppunct}\relax
\EndOfBibitem
\bibitem[Potzger \latin{et~al.}(2006)Potzger, Zhou, Reuther, M{\"{u}}cklich,
  Eichhorn, Schell, Skorupa, Helm, Fassbender, Herrmannsd{\"{o}}rfer, and {et
  al.}]{c18}
Potzger,~K.; Zhou,~S.; Reuther,~H.; M{\"{u}}cklich,~A.; Eichhorn,~F.;
  Schell,~N.; Skorupa,~W.; Helm,~M.; Fassbender,~J.; Herrmannsd{\"{o}}rfer,~T.;
  {et al.}, {Fe Implanted Ferromagnetic ZnO}. \emph{Appl. Phys. Lett.}
  \textbf{2006}, \emph{88}, 052508--3\relax
\mciteBstWouldAddEndPuncttrue
\mciteSetBstMidEndSepPunct{\mcitedefaultmidpunct}
{\mcitedefaultendpunct}{\mcitedefaultseppunct}\relax
\EndOfBibitem
\bibitem[McLeod \latin{et~al.}(2012)McLeod, Moewes, Zatsepin, Kurmaev, Wypych,
  Bobowska, Opasinska, and Cholakh]{mcleod_12}
McLeod,~J.~A.; Moewes,~A.; Zatsepin,~D.~A.; Kurmaev,~E.~Z.; Wypych,~A.;
  Bobowska,~I.; Opasinska,~A.; Cholakh,~S.~O. {Predicting the Band Gap of
  Ternary Oxides containing 3${d}^{10}$ and 3${d}^{0}$ Metals}. \emph{Phys.
  Rev. B} \textbf{2012}, \emph{86}, 195207--7\relax
\mciteBstWouldAddEndPuncttrue
\mciteSetBstMidEndSepPunct{\mcitedefaultmidpunct}
{\mcitedefaultendpunct}{\mcitedefaultseppunct}\relax
\EndOfBibitem
\bibitem[Mauchamp \latin{et~al.}(2009)Mauchamp, Jaouen, and
  Schattschneider]{mauchamp_09}
Mauchamp,~V.; Jaouen,~M.; Schattschneider,~P. {Core-hole Effect in the
  One-Particle Approximation Revisited from Density Functional Theory}.
  \emph{Phys. Rev. B} \textbf{2009}, \emph{79}, 235106--16\relax
\mciteBstWouldAddEndPuncttrue
\mciteSetBstMidEndSepPunct{\mcitedefaultmidpunct}
{\mcitedefaultendpunct}{\mcitedefaultseppunct}\relax
\EndOfBibitem
\bibitem[Jin \latin{et~al.}(2010)Jin, Chang, Boukhvalov, Zhang, Finkelstein,
  Xu, Zhou, Kurmaev, and Moewes]{c20}
Jin,~J.; Chang,~G.; Boukhvalov,~D.; Zhang,~X.; Finkelstein,~L.; Xu,~W.;
  Zhou,~Y.; Kurmaev,~E.; Moewes,~A. {Element-specific Electronic Structure of
  Mn Dopants and Ferromagnetism of (Zn,Mn)O Thin Films}. \emph{Thin Solid
  Films} \textbf{2010}, \emph{518}, 2825--2829\relax
\mciteBstWouldAddEndPuncttrue
\mciteSetBstMidEndSepPunct{\mcitedefaultmidpunct}
{\mcitedefaultendpunct}{\mcitedefaultseppunct}\relax
\EndOfBibitem
\bibitem[Mulliken(1955)]{mulliken_55}
Mulliken,~R.~S. {Electronic Population Analysis on LCAO-MO Molecular Wave
  Functions. I}. \emph{J. Chem. Phys.} \textbf{1955}, \emph{23},
  1833--1840\relax
\mciteBstWouldAddEndPuncttrue
\mciteSetBstMidEndSepPunct{\mcitedefaultmidpunct}
{\mcitedefaultendpunct}{\mcitedefaultseppunct}\relax
\EndOfBibitem
\bibitem[Mantovan \latin{et~al.}(2012)Mantovan, Gunnlaugsson, Naidoo,
  \'{O}lafsson, Johnston, Masenda, Molholt, Bharuth-Ram, Fanciulli, Gislason,
  and {et al.}]{c23}
Mantovan,~R.; Gunnlaugsson,~H.~P.; Naidoo,~D.; \'{O}lafsson,~S.; Johnston,~K.;
  Masenda,~H.; Molholt,~T.~E.; Bharuth-Ram,~K.; Fanciulli,~M.; Gislason,~H.~P.;
  {et al.}, {Fe Charge State Adjustment in ZnO upon Ion Implantation.} \emph{J.
  Phys.: Condens. Matter} \textbf{2012}, \emph{24}, 485801--5\relax
\mciteBstWouldAddEndPuncttrue
\mciteSetBstMidEndSepPunct{\mcitedefaultmidpunct}
{\mcitedefaultendpunct}{\mcitedefaultseppunct}\relax
\EndOfBibitem
\bibitem[Wang \latin{et~al.}(2009)Wang, Ding, Li, Song, and Wang]{wang_09}
Wang,~X.; Ding,~Y.; Li,~Z.; Song,~J.; Wang,~Z.~L. {Single-Crystal Mesoporous
  ZnO Thin Films Composed of Nanowalls}. \emph{J. Phys. Chem. C} \textbf{2009},
  \emph{113}, 1791--1794\relax
\mciteBstWouldAddEndPuncttrue
\mciteSetBstMidEndSepPunct{\mcitedefaultmidpunct}
{\mcitedefaultendpunct}{\mcitedefaultseppunct}\relax
\EndOfBibitem
\bibitem[Han \latin{et~al.}(2002)Han, Song, Yang, Park, Park, Jeong, and
  Rhie]{c24}
Han,~S.-J.; Song,~J.~W.; Yang,~C.-H.; Park,~S.~H.; Park,~J.-H.; Jeong,~Y.~H.;
  Rhie,~K.~W. {A Key to Room-Temperature Ferromagnetism in Fe-doped ZnO:Cu}.
  \emph{Appl. Phys. Lett.} \textbf{2002}, \emph{81}, 4212--4214\relax
\mciteBstWouldAddEndPuncttrue
\mciteSetBstMidEndSepPunct{\mcitedefaultmidpunct}
{\mcitedefaultendpunct}{\mcitedefaultseppunct}\relax
\EndOfBibitem
\bibitem[Wang \latin{et~al.}(2010)Wang, Su, Liu, Tian, Chang, Wang, Yin, and
  Yuan]{c26}
Wang,~Y.~Q.; Su,~L.; Liu,~L.; Tian,~Z.~M.; Chang,~T.~Q.; Wang,~Z.; Yin,~S.~Y.;
  Yuan,~S.~L. {Ferromagnetism in Fe-doped ZnO Bulk Samples with Additional Cu
  Doping}. \emph{Phys. Stat. Sol. (a)} \textbf{2010}, \emph{207},
  2553--2557\relax
\mciteBstWouldAddEndPuncttrue
\mciteSetBstMidEndSepPunct{\mcitedefaultmidpunct}
{\mcitedefaultendpunct}{\mcitedefaultseppunct}\relax
\EndOfBibitem
\bibitem[Zhou \latin{et~al.}(2008)Zhou, Potzger, Talut, Reuther, von Borany,
  Gr{\"{o}}tzschel, Skorupa, Helm, Fassbender, Volbers, and {et al.}]{c27}
Zhou,~S.; Potzger,~K.; Talut,~G.; Reuther,~H.; von Borany,~J.;
  Gr{\"{o}}tzschel,~R.; Skorupa,~W.; Helm,~M.; Fassbender,~J.; Volbers,~N.; {et
  al.}, {Fe-Implanted ZnO: Magnetic Precipitates versus Dilution}. \emph{J.
  Appl. Phys.} \textbf{2008}, \emph{103}, 023902--14\relax
\mciteBstWouldAddEndPuncttrue
\mciteSetBstMidEndSepPunct{\mcitedefaultmidpunct}
{\mcitedefaultendpunct}{\mcitedefaultseppunct}\relax
\EndOfBibitem
\bibitem[Yafet(1987)]{yafet_87}
Yafet,~Y. {Ruderman-Kittel-Kasuya-Yosida Range Function of a One-Dimensional
  Free-Electron Gas}. \emph{Phys. Rev. B} \textbf{1987}, \emph{36},
  3948--3949\relax
\mciteBstWouldAddEndPuncttrue
\mciteSetBstMidEndSepPunct{\mcitedefaultmidpunct}
{\mcitedefaultendpunct}{\mcitedefaultseppunct}\relax
\EndOfBibitem
\end{mcitethebibliography}
\end{document}